\documentclass{article}
\pdfoutput=1

\usepackage{amssymb,amsfonts,amsmath,stmaryrd}
\usepackage{cite,enumerate,float,indentfirst}
\usepackage{color}
\usepackage{tikz}
\usepackage{array}
\usepackage{ytableau}

\def\be{\begin{eqnarray}}
\def\ee{\end{eqnarray}}
\def\nn{\nonumber}
\def\p{\partial}

\def\Tr{{\rm Tr}\,}

\def\p{\partial}

\def\Tr{{\rm Tr}\,}

\definecolor{red}{rgb}{1,0,0}
\definecolor{orange}{rgb}{1,0.5,0}
\definecolor{violet}{rgb}{0.7,0,1}
	\definecolor{grannysmithapple}{rgb}{0.66, 0.89, 0.63}

\hoffset= - 1.0in         % switch off for draft style
\begin{document}

\title{\vspace{1.5cm}\bf
Two $\beta$-ensemble realization of $\beta$-deformed WLZZ models
}

\author{
A. Mironov$^{b,c,d,}$\footnote{mironov@lpi.ru,mironov@itep.ru},
A. Oreshina$^{a,c,}$\footnote{oreshina.aa@phystech.edu},
A. Popolitov$^{a,c,d,}$\footnote{popolit@gmail.com}
}

\date{ }

\maketitle

\vspace{-6.5cm}

\begin{center}
\hfill FIAN/TD-05/24\\
\hfill IITP/TH-07/24\\
\hfill ITEP/TH-08/24\\
\hfill MIPT/TH-06/24
\end{center}

\vspace{4.5cm}

\begin{center}
$^a$ {\small {\it MIPT, Dolgoprudny, 141701, Russia}}\\
$^b$ {\small {\it Lebedev Physics Institute, Moscow 119991, Russia}}\\
$^c$ {\small {\it NRC ``Kurchatov Institute", 123182, Moscow, Russia}}\\
$^d$ {\small {\it Institute for Information Transmission Problems, Moscow 127994, Russia}}
\end{center}

\vspace{.1cm}

\begin{abstract}
We consider a two $\beta$-ensemble realization of the series of $\beta$-deformed WLZZ matrix models. We demonstrate that such a realization involves $\beta$-deformed Harish-Chandra-Itzykson-Zuber integrals, one of them providing a coupling to the external field. We also construct Ward identities in the corresponding two $\beta$-ensemble model, which requires a set of identities for partition function of the one $\beta$-ensemble in the external field, and a set of identities for the $\beta$-deformed Itzykson-Zuber integral. These both sets of identities are formulated in terms of the Dunkl operators.
\end{abstract}

\section{Introduction}

Originally, matrix models were defined as integrals over matrices \cite{Dyson1,Dyson2,Dyson3,Dyson4,Mehta}. However, in the course of further development, there was realized a series of typical properties of matrix models that allowed one to use various definitions of matrix models not immediately dealing with matrix integrals. For instance, in the case of invariant integrand, one can integrate over angular variables so that there remain only $N$ integrations over the eigenvalues of the matrix. Deformations of these eigenvalue integrals can sometimes no longer be presented as a matrix integral, but is still often referred to as a matrix model. A typical examples are given by the $\beta$-ensembles \cite{Dyson1,Dyson2,Dyson3,Dyson4,Mehta,beta1,beta2,beta3,beta4,Ey1} and by the conformal matrix models \cite{confmamo1,confmamo2,confmamo3,confmamo4}. Another example is provided by $W$-representation of the matrix model partition function \cite{Wrep1,Wrep2,Wrep3,Wrep4,Max1,MMMR,MMM} (see also similar realizations in \cite{Giv,Ok,AMM1,AMM2,AMM3,AMM4,BM1,BM2,BM3}), i.e. reproducing the matrix model partition function by action on a trivial function (unit or linear exponential) with an exponential of a differential operator: one can deform such a representation, and the model is still often called matrix model even though no explicit matrix integral representation may be known.

As for the typical properties of matrix models, the two main features have long been known \cite{UFN31,UFN32,UFN33,UFN34,UFN35,UFN36} and include integrability \cite{Mamoint1,Mamoint2,Mamoint3,Mamoint4,Mamoint5,GKM} and an infinite set of Ward identities that is satisfied by the matrix model partition function \cite{Vir1,Vir2,Vir3,Vir4}. Recently, a new typical feature was discovered and amply discussed: it is the so called superintegrability originally defined in \cite{IMM,MM} (based on the phenomenon earlier observed in \cite{DiF1}--\cite{Pop}, see also some preliminary results in \cite{Kaz}--\cite{MKR} and later progress in \cite{MMten}--\cite{MO},\cite{Max1},\cite{MMssi}--\cite{MMPS2}). The property of superintegrability has inspired in \cite{WLZZ1,WLZZ2} constructing two infinite series of models that are given by their $W$-representations and are called WLZZ models. Note that these models include, as particular cases, the Gaussian Hermitian and complex matrix models.
Matrix representation interpolating between all these models was found later \cite{Ch1,Ch2} and turned out to be given by a two-matrix model (generally, in external field), which reduces to one-matrix models in the very particular cases.

Speaking more concretely, one can distinguish the four essential points related to the WLZZ models:

\begin{itemize}
\item First of all, the authors of \cite{WLZZ1,WLZZ2} proposed two infinite sets of differential operators in variables $p_k$, $\hat H_n(p)$ and $\hat H_{-n}(p)$, $n\in\mathbb{N}$ such that they generate matrix model partition functions (i.e. realize $W$-representations)\footnote{In \cite{WLZZ1,WLZZ2,MMsc,Ch1,Ch2}, the notation for these partition functions was opposite: $Z_n\leftrightarrow Z_{-n}$. Here we use the notation that match definitions in papers \cite{MMMP1,MMMP2}. }:
\be\label{1}
Z_n(p)=e^{\hat H_n(p)\over n}\cdot 1=\sum_R\xi_R(N)S_R\{p_k=\delta_{k,n}\}S_R\{p_k\}
\ee
and
\be\label{2}
Z_{-n}(g,p)=e^{\hat H_{-n}(p)\over n}\cdot e^{\sum_k{g_kp_k\over k}}
=\sum_{R,Q}{\xi_R(N)\over \xi_Q(N)}S_{R/Q}\{p_k=\delta_{k,n}\}S_R\{g_k\}S_Q\{p_k\}
\ee
where $N$ is a parameter that may be identified with the size of matrix in the matrix model, the sums run over all partitions (Young diagrams) $R$, $\xi_R(N)$ is a product over the boxes $(i,j)$ of the Young diagram:
$\xi_R(N) = \prod_{(i,j)\in R} (N+i-j)$, the Schur functions $S_R\{p_k\}$ are labeled by the Young diagram $R$ and are graded polynomials of $p_k$, and $S_{R/Q}\{p_k\}$ are the skew Schur functions \cite{Mac}.

\item The operators $\hat H_{\pm n}$ form two commutative families, and generate an interpolating two-matrix model \cite{Ch1,Ch2}:
\be\label{im}
Z(N;\bar p,p,g)=e^{\sum_n{\bar p_n\hat H_{-n}(p)\over n}}\cdot e^{\sum_k{g_kp_k\over k}}
=\sum_R{\xi_R(N)\over \xi_Q(N)}S_{R/Q}\{\bar p_k\}S_R\{g_k\}S_Q\{p_k\}=\nn\\
=\int\int_{N\times N}[dXdY]\exp\left(-\Tr XY+\Tr Y\Lambda+\sum_k {g_k\over k}\Tr X^k+\sum_k{\bar p_k\over k}\Tr Y^k\right)
\ee
with $p_k=\Tr\Lambda^k$.  Here the integral is understood as integration of a power series in $g_k$, $\bar p_k$ and $\Tr\Lambda^k$, and $X$ are Hermitian matrices, while $Y$ are anti-Hermitian ones. The measure is normalized in such a way that $Z(N;0,0,0)=1$.

Now one immediately obtains
\be
Z_n(p)=Z(N;p,0,\delta_{k,n})\nn\\
Z_{-n}(g,p)=Z(N;\delta_{k,n},p,g)
\ee

\item It turns out \cite{MMCal,MMMP1} that the operators $\hat H_{\pm n}$ are elements of a commutative subalgebra of the $W_{1+\infty}$-algebra \cite{Pope1,Pope2,Pope3,Pope4,FKN2,BK1,BK2,KR1,KR2,KR3,Awata,Miki}. One can consider more commutative subalgebras associated with so called ``integer rays" \cite{MMMP1}. Their elements $\hat H_{\pm n}^{(m)}$ give rise to more WLZZ models \cite{Ch1,Ch2}, which, however, do not have a matrix model representation so far\footnote{One can consider more general commutative subalgebras in $W_{1+\infty}$, those associated with ``rational rays", however, their $\beta$-deformation, which is the point of our interest in this paper, meets some problems \cite{MMMP2}.}

\item As any matrix model, the partition function $Z(N;\bar p,p,g)$ satisfies an infinite set of constraints, the Ward identities. They are generally very involved even for $Z_{-n}(g,p)$ (see \cite{MMsc}, where the case of $Z_{-2}(g,p)$ is considered in the very detail), however, the set of constraints for $Z_n(p)$ is treatable. That is, the partition function $Z_n(p)$ satisfies the infinite set of constraints \cite{MMM91,GKMU}
\be\label{Wmcon}
\widetilde {W}_{k}^{(-,n)}Z_{n}(p)=(k+n){\p Z_{n}(p)\over\p p_{k+n}}, \ \ \ \ \ k\ge -n+1
\ee
where $\widetilde {W}_{k}^{(-,n)}$ are the generators of the $\widetilde W$-algebra introduced in \cite{MMM91,GKMU}. Moreover, the $W$-representation is generated by this $\widetilde W$-algebra \cite{MMsc}:
\be
\hat H_n(p)=\sum_{k}p_k\widetilde {W}_{k-n}^{(-,m)}
\ee
and similarly for $H_{-n}$ \cite{MMsc} and even for $\hat H_{\pm n}^{(m)}$ \cite{DMP}, but this latter requires certain natural generalization of the $\widetilde W$-algebra.
\end{itemize}

Now note that all the four points admit a $\beta$-deformation. Indeed, the $W$-representation was $\beta$-deformed in \cite{WLZZ2,Ch2,MO}\footnote{In fact, the construction admits even a $(q,t)$-deformation, as was demonstrated in \cite{Ch3}.}. The commutative subalgebras of the $W_{1+\infty}$ algebra are substituted \cite{MMMP2} by those of the affine Yangian $Y( \Hat{gl}_{\infty} )$ \cite{Yangian1,Yangian2} under the $\beta$-deformation. As for the matrix model representation and for the corresponding infinite set of the Ward identities, this is {\bf the main goal of the present paper} to present their $\beta$-deformation.

That is, we demonstrate that the matrix model integral (\ref{im}) is replaced under the $\beta$-deformation by the multiple integral
    \begin{equation} \label{PartFuncLambda}
    \boxed{
Z^{(\beta)}(N;\bar p,p,g)=\int[dxdy]  \Delta^{2\beta} (x) \Delta^{2\beta} (y) I^{\beta}(x,-y)  I^{\beta}(\lambda,y) \exp \left[ \sum_{k \geq 1} \frac{g_k}{k} \sum_{j=1}^{N} x_j^k +  \sum_{k \geq 1} \frac{\bar p_k}{k}\sum_{j=1}^{N} y_j^m \right]
    }
    \end{equation}
i.e. by two $\beta$-ensembles $x_i$ and $y_i$ instead of two matrices $X$ and $Y$, and $p_k=\sum_j\lambda_j^k$. Here the integrals over $x_j$ run over the real axis, and integrals over $y_j$ run over the imaginary ones. $I^{\beta}(x,y)$ is the $\beta$-deformed Harish-Chandra-Itzykson-Zuber ($\beta$-HCIZ) integral, and the measure is again normalized in such a way that $Z_\beta(N;0,0,0)=1$.

Our main claim in the paper is that this integral (\ref{PartFuncLambda}) is equal to the $\beta$-deformed WLZZ model \cite{WLZZ2,Ch1,MMMP1}
\be\label{WLZZ}
\boxed{
Z^{(\beta)}(N;\bar p,p,g)=\sum_{R,P}{\xi_R^\beta(N)\over \xi_P^\beta(N)}{J_{R/P}\{\bar p_k\}J_R\{g_k\}J_P(\lambda)\over ||J_R||}=e^{\sum_n{\beta^{1-n}\over n}\bar p_n\hat H_{-n}(p)}\cdot e^{\beta\sum_k{g_kp_k\over k}}
}
\ee
Where $J_R\{p_k\}$ denotes the Jack polynomial, which is the $\beta$-deformation of the Schur polynomial $S_R\{p_k\}$, and, hereafter, we use the notation $J_R(\lambda)=J_R\{\sum_j\lambda_j^k\}$.

Furthermore, we construct an infinite set of Ward identities satisfied by the partition function $Z_n^\beta(p)=Z_\beta(N;\delta_{k,n},0,p)$ and find a $\beta$-deformed counterpart of the $\widetilde W$-algebra. It is interesting that this latter is constructed with the Dunkl operators \cite{Dunkl} substituting the matrix derivatives, much similar to the observation of \cite{MMCal}. At the core of these Ward identities is a rather striking family of identities (\ref{WIIZ}) for the $\beta$-HCIZ integral. These identities are novel and non-trivially generalize the earlier known fact that the $\beta$-HCIZ integral is an eigenfunction of the rational Calogero model \cite{Ey,MMMP1}.

This paper is organized as follows. In section 2, we explain what has been known about the $\beta$-deformation of the $W$-representation of the WLZZ models so far, and discuss their various properties. In section 3, we first discuss properties of the HCIZ and $\beta$-HCIZ integrals and their convenient representations, then present the two $\beta$-ensemble representation (\ref{PartFuncLambda}) at all $\lambda_j=0$, and, at last, extend it to the general representation (\ref{PartFuncLambda}). In section 4, we discuss Ward identities for the partition function $Z_n^\beta(p)=Z_\beta(N;\delta_{k,n},0,p)$, which involve a non-trivial identity for the $\beta$-HCIZ integral (\ref{WIIZ}). These Ward identities are associated with a $\beta$-deformation of the $\widetilde W$-algebra. Using the Ward identity allows us to prove the main claim of the paper. Section 5 contains some concluding remarks, and the Appendix, various illustrations of the basic formula (\ref{main}) used in the derivation of the Ward identities.

\paragraph{Notation.} Throughout the paper, we use the notation $\int [dxdy]$ for $\int\ldots \int  \prod_{i=1}^{N} dx_i dy_i$, and $[dXdY]$ for the integration measure on the Hermitian matrices (or $[dU]$, in section 3, for the Haar measure on the unitary group). We also use the notation $S_R(x)$, $J_R(x)$, etc. for symmetric functions of variables $x_j$, and $S_R\{p_k\}$, $J_{R}\{p_k\}$, etc. for graded polynomials that are the functions of power sums $p_k=\sum_jx_j^k$.

\section{$\beta$-deformed WLZZ models}

\subsection{$W$-representation of the WLZZ models}

We start with a description of the $\beta$-deformed WLZZ models \cite{WLZZ2,Ch1,MMMP1}. As we explained in the Introduction, there are two sets of models generated by two commutative families of Hamiltonians. In order to construct these Hamiltonians, we define the cut-and-join operator
\begin{align}\label{W0b}
\hat W_0 :=  & \ \frac{1}{2}\sum_{a,b=1} \left(abp_{a+b}\frac{\p^2}{\p p_a\p p_b} + \beta (a+b)p_ap_b\frac{\p}{\p p_{a+b}}\right)
+ \beta u_\beta\sum_{a=1} ap_a\frac{\p}{\p p_a}+{\beta u_\beta^3\over 6}
+{1-\beta\over 2}\sum_aa(a-1)p_a{\p\over\p p_a}= \\ \notag
= & \ {\beta^2\over 6}\sum_{a,b,c\in\mathbb{Z}}^{a+b+c=0}:p_ap_bp_c:+ {\beta(1 - \beta)\over 4}\sum_{a,b\in\mathbb{Z}}^{a+b=0} \left(|a| - 1\right) :p_ap_{b}:
\end{align}
where $p_0=u_\beta=N+(\beta-1)/(2\beta)$, $p_{-k}=\beta^{-1}k{\p\over\p p_k}$ and the normal ordering means all derivatives put to the right.

We also define operators that generate the families of Hamiltonians:
\begin{align}
   \hat E_1 = & \ [\hat W_0, p_1] = \frac{\beta}{2} \sum_{a+b=1} :p_{a} p_{b}:
    \\ \notag
   \hat E_2 = & \ [\hat W_0,\hat E_1] = \frac{\beta^2}{3} \sum_{a+b+c=1}^\infty :p_a p_b p_c:
   + \beta (1 - \beta) \sum_{k=0}^\infty k p_{k+1} p_{-k}
\end{align}
and
\begin{align}
\hat F_1&= \beta \sum_{b=0} p_b p_{-b-1} \\ \notag
\hat{F}_2 & = [\hat F_1,\hat W_0] =
\frac{\beta^2}{3} \sum_{a+b+c = -1} : p_a p_b p_c : + \beta(1-\beta) \sum_b b \cdot p_b p_{-b-1}
\end{align}
Then, the commutative families are recursively defined by
\be
\hat H_{n+1}={1\over n}[\hat E_2,\hat H_n]
\ee
with the initial condition $\hat H_1=\hat E_1$,
and
\be
\hat H_{-n-1}={1\over n}[\hat H_{-n},\hat F_2]
\ee
with the initial condition $\hat H_{-1}=\hat F_1$.

Note that the both series of commutative operators $\hat H_{n}$ and $\hat H_{-n}$ are associated with Hamiltonians of the rational Calogero model \cite{MMCal,MMMP1}.

Now we define
\be\label{1b}
Z_n^{(\beta)}(p)=e^{{\beta^{1-n}\over n}\hat H_n(p)}\cdot 1
\ee
and
\be\label{2b}
Z_{-n}^{(\beta)}(g,p)=e^{{\beta^{1-n}\over n}\hat H_{-n}(p)}\cdot e^{\beta\sum_k{g_kp_k\over k}}
\ee
and the interpolating model
\be\label{3b}
Z^{(\beta)}(N;\bar p,p,g)=e^{\sum_n{\beta^{1-n}\over n}\bar p_n\hat H_{-n}(p)}\cdot e^{\beta\sum_k{g_kp_k\over k}}
\ee
These partition functions are $\beta$-deformations of sums (\ref{1})-(\ref{im}) and can be presented as sums over partitions of products of the Jack and skew Jack polynomials instead of the Schur ones. Hence, in the next subsection we describe their properties.

\subsection{Jack polynomials}

Now we consider properties of polynomials $J_R\{p_k\}$ and $J_{R/Q}\{p_k\}$, which are the Jack and skew Jack polynomials accordingly \cite{Mac}, and which realize a proper $\beta$-deformation of the Schur and skew Schur polynomials.

First of all note that the commuting operators introduced in the previous subsection can be simply described in terms of an operator
$\hat {\cal O}^\beta_N$ with the property
\be
    \hat {\cal O}^\beta(N) \cdot J_R\{p_k\} =  \xi_R^\beta(N)\  J_R\{p_k\}\nn\\
    \xi_R^\beta(N):=\prod_{i,j\in N}(N+\beta^{-1}(j-1)-i+1)
\ee
Construction of this operator is discussed in \cite{Ch2,MMMP1,MMMP2}, here we do not need its manifest form. What is important, one can realize \cite{Ch2,MMMP1,MMMP2}
\be\label{Wk}
  \hat H_{-k}= \Big(\hat {\cal O}^\beta(N)\Big)^{-1}\left( {k\over\beta}\dfrac{\partial}{\partial p_k}\right) \hat {\cal O}^\beta(N)\nn\\
  \hat H_k= \hat {\cal O}^\beta(N)p_k \Big(\hat {\cal O}^\beta(N)\Big)^{-1}
\ee

We also need an orthogonality relation for the Jack polynomials,
\be\label{sJ}
J_Q\left\{{k\over\beta}{\p\over\p p_k}\right\}\ J_R\{p_k\}=||J_Q||\ J_{R/Q}\{p_k\}
\ee
where $||J_Q||$ is the norm square of the Jack polynomial,
\be
||J_R||:={\overline{G}^\beta_{R^\vee R}(0)\over G^\beta_{RR^\vee}(0)}\beta^{|R|}\ \ \ \ \ \ \
G_{R'R''}^\beta(x):=\prod_{(i,j)\in R'}\Big(x+R'_i-j+\beta(R''_j- i+1)\Big)
\ee
with the bar over the functions denoting the substitution $\beta\to\beta^{-1}$.

Now let us use the identity (\ref{sJ}) in order to obtain
\be
J_R\left\{\hat H_{-k}\right\}J_Q\{p_k\}= \xi_Q^\beta(N)\Big(\hat {\cal O}^\beta(N)\Big)^{-1}J_R\left\{{k\over\beta}{\p\over\p p_k}\right\}J_Q\{p_k\}=\xi_Q^\beta(N)||J_Q||\Big(\hat {\cal O}^\beta(N)\Big)^{-1}J_{Q/R}\{p_k\}=\nn\\
=\xi_Q^\beta(N)||J_Q||\Big(\hat {\cal O}^\beta(N)\Big)^{-1}\sum_P\phantom{.}^{\beta^{-1}} N^{Q^\vee}_{R^\vee P^\vee}J_P\{p_k\}=
\sum_P\phantom{.}^{\beta^{-1}} N^{Q^\vee}_{R^\vee P^\vee}||J_Q||{\xi_Q^\beta(N)\over\xi_P^\beta(N)}J_P\{p_k\}
\ee
where $\phantom{.}^\beta N^Q_{RP}$ are the Littlewood-Richardson coefficients, and we used that
\be
J_{R/P}=\sum_Q\phantom{.}^{\beta^{-1}} N^{R^\vee}_{Q^\vee P^\vee}\ J_Q
\ee
Note that
\be\label{N}
\phantom{.}^{\beta^{-1}} N^{R^\vee}_{Q^\vee P^\vee}=\phantom{.}^\beta N^{R}_{Q P}{||J_R||\over ||J_Q||\ ||J_P||}
\ee
In particular,
\be\label{OP}
J_R\left\{\hat H_{-k}\right\}J_Q\{p_k\}\Big|_{p_k=0}=\xi_R^\beta(N)||J_Q||\delta_{R,Q}
\ee

Now note that the partition functions of the previous subsection can be rewritten in terms of sums over the Jack polynomials \cite{WLZZ2,MO,Ch1,Ch2}:
\be\label{bZ}
Z_n^{(\beta)}(p)=\sum_R\xi_R^{(\beta)}(N){J_R\{p_k=\delta_{k,n}\}J_R\{p_k\}\over ||J_R||}\nn\\
Z_{-n}^{(\beta)}(g,p)=\sum_R{\xi^{(\beta)}_R(N)\over \xi^{(\beta)}_Q(N)}
{J_{R/Q}\{p_k=\delta_{k,n}\}J_R\{g_k\}J_Q\{p_k\}\over ||J_R||}\nn\\
Z^{(\beta)}(N;\bar p,p,g)=\sum_R{\xi^{(\beta)}_R(N)\over \xi^{(\beta)}_Q(N)}{J_{R/Q}\{\bar p_k\}J_R\{g_k\}J_Q\{p_k\}
\over ||J_R||}
\ee

\subsection{Towards $\beta$-ensemble realization}

Now we introduce a $\beta$-deformation of the matrix model (\ref{im}) in the following way:
\be\label{imJ}
Z_{mm}^\beta(N;\bar p,p,g)=
\int[dxdy]  \mu(x,y,\lambda) \exp \left[ \sum_{k \geq 1} \frac{g_k}{k} \sum_{j=1}^{N} x_j^k +  \sum_{k \geq 1} \frac{\bar p_k}{k}\sum_{j=1}^{N} y_j^m \right]
\ee
where the $\beta$-deformed integration measure $\mu(x,y,\lambda)$ is {\it defined} as
\be\label{meas}
\boxed{
\int [dxdy]  \mu(x,y,\lambda) J_R(x)J_Q(y)
=J_R\left\{\hat H_{-k}\right\}J_Q(y)I^\beta(y,\lambda)\Big|_{y=0}
}
\ee
The integrals over $x_j$ run along the real axis, and those over $y_j$, over the imaginary one.

 In order to evaluate the matrix integral (\ref{imJ}), we use the manifest expression for the $\beta$-HCIZ integral \cite[sec.2]{MMS} (see a discussion in the next section):
\be\label{IZb}
I^\beta(y,\lambda)=\sum_P{1\over\xi_P^\beta}{J_P(y)J_P(\lambda)
\over ||J_P||}
\ee
Then, using the Cauchy identity for the Jack polynomials
\be
\sum_R{J_R\{p_k\}J_R\{p'_k\}\over ||J_R||}=\exp\left(\beta\sum_k{p_kp'_k\over k}\right)
\ee
and, using (\ref{OP}), one immediately gets that
\be\label{imJf}
\boxed{
Z_{mm}^\beta(N;\bar p,p,g)
=\sum_{R,Q,P}{\xi_R^\beta(N)\over \xi_P^\beta(N)}\phantom{.}^\beta N^R_{QP}{J_R\{g_k\}J_Q\{\bar p_k\}J_P(\lambda)\over ||J_P||\ ||J_Q||}\stackrel{(\ref{N})}{=}\sum_{R,P}{\xi_R^\beta(N)\over \xi_P^\beta(N)}{J_{R/P}\{\bar p_k\}J_R\{g_k\}J_P(\lambda)\over ||J_R||}
}
\ee
i.e. $Z_{mm}^\beta(N;\bar p,p,g)=Z^\beta(N;\bar p,p,g)$ from (\ref{bZ}).

\section{Two $\beta$-ensemble realization of the WLZZ models}

\subsection{$\beta$-HCIZ integral}

In this section, we obtain the measure $\mu(x,y,\lambda)$ that provides the necessary property (\ref{meas}). Since it involves the $\beta$-HCIZ integrals, we discuss here what is this latter.

\paragraph{The HCIZ integral.} The usual HCIZ integral \cite{HC,IZ} is the matrix integral over the unitary group $U(N)$:
    \begin{equation}\label{IZInt}
        I_N(X,Y) = \int [dU] \exp \left[ \Tr(U^{\dagger} X U Y) \right]
    \end{equation}
which we normalize to be $I(0,0)=1$. The answer for this integral is a function of eigenvalues $x_j$, $y_j$, $j=1\ldots N$ of matrices $X$ and $Y$ accordingly, and has two equivalent representations:
\begin{itemize}
\item in terms of Schur functions \cite{Kazak,Bal,Mor}:
    \begin{equation}\label{IZSchur}
        I_N(X,Y) = \sum_{R} {1\over\xi_R(N)}S_R(x) S_R(y)
    \end{equation}
where we denote with $S_R(x)=S_R(X)$ a symmetric function of variables $x_j$ such that $S_R\{p_k\}$ is the Schur functions of power sums of these variables: $p_k=\sum_jx_j^k$, with the sum here running over partitions with number of parts\footnote{Otherwise the Schur functions vanish, and so does $\xi_R(N)$, and cancellation of the integrand at $l_R> N$ can be automatically achieved by a proper limiting procedure.} $l_R\le N$;
\item as a determinant (Harish-Chandra form) \cite{HC}:
        \begin{equation}\label{IZdet}
            I_N(X,Y) =\frac{\det_{j,k} \left[ e^{x_j y_k} \right]}{ \Delta(x) \Delta(y) }
        \end{equation}
where $\Delta(x)=\prod_{j<k}(x_j-x_k)$ is the Vandermonde determinant.
\end{itemize}

\paragraph{$\beta$-HCIZ integral.}

\begin{itemize}
\item The first representation (\ref{IZSchur}) is naturally $\beta$-deformed by replacing the Schur functions with the Jack polynomials \cite{MMS}:
\begin{equation}\label{IZJack}
        I_N^\beta(x,y) = \sum_{R} {1\over \xi_R^{(\beta)}(N)}{J_R(x) J_R(y)\over||J_R||}
    \end{equation}
At $\beta=1/2$ and $\beta=2$ this expression is nothing but the integral \eqref{IZInt} over orthogonal and symplectic matrices correspondingly, which, indeed, reduces to (\ref{IZJack}). Again, the sum runs over partitions with number of parts $l_R\le N$. From now on, we do not specially stipulate it.
\end{itemize}

However, at generic $\beta$ an integral representation is not available. Instead, there is a recurrent definition of the integral in $N$ \cite{Ey}:
\be\label{bIZd}
I^\beta_N(x,y)\sim {e^{x_N\sum_jy_j}\over \prod_{j-1}^{N-1}(x_j-x_N)^{2\beta-1}}\int d\lambda_1\ldots d\lambda_{N-1}
{\Delta(\lambda)^{2\beta}e^{x_N\sum_j\lambda_j}\over\prod_{k=1}^N\prod_{j=1}^{N-1}(\lambda_j-y_k)^\beta}\ I^\beta_{N-1}(x_{N-1},\lambda)
\ee
with the initial condition of the recursion $I_1^\beta(x,y)=\exp\Big(\beta xy\Big)$. However, this expression is ambiguous because of different possibilities of choosing the integration contours (they are such that the integral is convergent, and they surround all points $y_j$). Since the answer as a function of $x_j$ and $y_j$ sometimes has branching (depending on values of $\beta$), different choices of the integration contours may correspond to different branches. Though it may only change the overall coefficient and seem relatively innocuous, this is not the case: since we are going to further integrate $I_N^\beta(X,Y)$ over all $x_j$ and $y_j$, the choice of integration contours/branch essentially influences the final result. The safe and unambiguous answers for the integrals are only at $\beta \in \mathbb{N}$. However, various differential equations do not depend on possible ambiguities and, hence, are valid at any $\beta$. This is what we are going to use in this paper: to establish various connections involving integrals at $\beta \in \mathbb{N}$, and, involving differential equations, at arbitrary $\beta$.

\begin{itemize}
\item Thus, in the case of $\beta \in \mathbb{N}$ following \cite{Brezin,Ey}, one can evaluate the integral (\ref{bIZd}) by the formula
        \begin{equation}\label{iizrec}
            I_{N}^\beta(x,y) = \sum_{\sigma} \frac{e^{\sum_{j=1}^{N} x_j y_{\sigma(j)}}}{\Delta(x)^{2\beta}\Delta(y_{\sigma})^{2\beta}} \Tilde{I}_N^{\beta} (x,y_{\sigma})
        \end{equation}
where the sum runs over permutations of variables $y_i$, and $\Tilde{I}_{N}^\beta(X,Y)$ is calculated recurrently:
        \begin{equation}
            \Tilde{I}_N^{\beta} (x,y) = \prod_{i=1}^{N-1} (x_j-x_{N}) (y_{N}-y_j)^{\beta} \left( \frac{\partial}{\partial \lambda_j} \right)^{\beta-1} e^{(x_j-x_{N}) \lambda_j} \prod_{k \neq j} ^N \frac{(y_k-y_j)^{\beta}}{(y_k - y_j - \lambda_j)^{\beta}} \Tilde{I}_{N-1}^{\beta} (x,y+\lambda)|_{\lambda=0}
        \end{equation}
 with the initial data $\Tilde{I}_1^{\beta} (x,y) = 1$.
 \end{itemize}

Let us note that the $\beta$-HCIZ integral provides eigenfunctions of the rational Calogero system \cite{Cal1,Cal2}: it was proved both for formula (\ref{IZJack}) in \cite[sec.12]{MMMP1} and for integral formulas in \cite{Ey}. This proves the equivalence of two representations for the $\beta$-HCIZ integral.

In fact, the Calogero Hamiltonians coincide with $\hat H_{-n}$ (or with $\hat H_n$, depending on the choice of coordinates) expressed through $x_j$ instead of $p_k=\sum_jx_j^k$ (see details in \cite{MMCal}) up to a simple conjugation, and the quadratic Hamiltonian is
        \be \label{HIeyg}
           \hat H_{-2}\cdot I^{\beta} (x,y)&=& \left(\sum_{j} y_j^2\right)\cdot I^{\beta} (x,y)\nn\\
\hat H_{-2}&=&\sum_{j} {\p^2\over\p x_j^2} + \beta \sum_{j,k;k\ne j} \frac{1}{x_j- x_k }\left({\p\over\p x_j}
-{\p\over\p x_k}\right)
\ee
This Hamiltonian is related with the standard rational Calogero Hamiltonian,
\be
\hat H_C=\sum_{j} {\p^2\over\p x_j^2}-\beta(\beta-1)\sum_{j,k;j\ne k}{1\over (x_j-x_k)^2}
\ee
by the transform $\hat H_C=\Delta(x)^\beta\cdot\hat H_{-2}\cdot\Delta(x)^{-\beta}$.

\subsection{Two $\beta$-ensemble realization at all $\lambda_j=0$}

As we explained in sec.2.3, in order to construct a two $\beta$-ensemble realization of the WLZZ models, one has to find a measure in the multiple integral (\ref{imJ}) that satisfies the equation
\be
\int [dxdy]  \mu(x,y,\lambda) J_R(x)J_Q(y)
=J_R\left\{\hat H_{-k}\right\}J_Q(y)I^\beta(y,\lambda)\Big|_{y=0}
\ee
Let us first consider the case of all $\lambda_j=0$, i.e. we look for the measure such that
\be\label{m1}
\int [dxdy]  \mu(x,y,0) J_R(x)J_Q(y)
=J_R\left\{\hat H_{-k}\right\}J_Q(y)\Big|_{y=0}\stackrel{(\ref{OP})}{=}\xi_R^\beta(N)||J_Q||\delta_{R,Q}
\ee
We claim that this property is provided by the measure
\be\label{me1}
\boxed{
\mu(x,y,0)=\Delta^{2\beta} (x) \Delta^{2\beta} (y) I^{\beta}(x,-y)
}
\ee
Note that, in order to check property (\ref{m1}) with this measure, one can not use the Jack polynomial expansion (\ref{IZJack}): the integrals of any concrete term in the sum over partitions diverges. Hence, one has to use formula (\ref{iizrec}), which is valid only at $\beta\in\mathbb{N}$. In this case each term comes with an exponential, and one can use the standard Fourier theory formula
\be\label{bas}
\left.\int dxdyf(x)g(y)e^{-xy}=f\left({\partial\over\partial x}\right)g(x)\right|_{x=0}
\ee
where the $x$-integral goes over the real axis, and the $y$-integral runs over the imaginary one.

We checked (\ref{m1}) with measure (\ref{me1}) for a few first values of $N$ and $\beta\in\mathbb{N}$, and confirmed it is correct.

There is another way to check this measure, which is suitable at arbitrary $\beta$ but is less immediate: one can find Ward identities for the multiple integral and construct the $W$-representation following \cite{MMMR}. We demonstrate in the next section how this works, and explain that the $W$-representation obtained coincides with (\ref{1b}).

\subsection{Generic two $\beta$-ensemble realization\label{gen}}

Now we claim that the full measure is
\be\label{me2}
\boxed{
\mu(x,y,\lambda)=\Delta^{2\beta} (x) \Delta^{2\beta} (y) I^{\beta}(x,-y)I^{\beta}(\lambda,y)
}
\ee
This can be demonstrated using only the known measure at all $\lambda_j=0$ (\ref{me1}), which guarantees formula (\ref{m1}).
Indeed,
\be
\int [dxdy]\Delta^{2\beta} (x) \Delta^{2\beta} (y) I^{\beta}(x,-y)I^{\beta}(\lambda,y) J_R(x)J_Q(y)\stackrel{(\ref{IZJack})}{=}\nn\\
=\int [dxdy] \Delta^{2\beta} (x) \Delta^{2\beta} (y) I^{\beta}(x,-y)I^{\beta}(\lambda,y) J_R(x)J_Q(y)\sum_P{1\over \xi_P^{(\beta)}(N)}{J_P(\lambda) J_P(y)\over||J_P||}=\nn\\
=\int [dxdy] \Delta^{2\beta} (x) \Delta^{2\beta} (y) I^{\beta}(x,-y)I^{\beta}(\lambda,y) J_R(x)
\sum_{P,T}\phantom{.}^\beta N^{T}_{Q P}{1\over \xi_P^{(\beta)}(N)}{J_P(\lambda)\over||J_P||}
J_T(y)\stackrel{(\ref{m1})}{=}\nn\\
=\sum_{P,T}\phantom{.}^\beta N^{T}_{Q P}{1\over \xi_P^{(\beta)}(N)}{J_P(\lambda)\over||J_P||}
J_R\left\{\hat H_{-k}\right\}J_T(y)\Big|_{y=0}\stackrel{(\ref{IZJack})}{=}
J_R\left\{\hat H_{-k}\right\}J_Q(y)I^\beta(y,\lambda)\Big|_{y=0}
\ee
This completes our proof of the main claim in this paper. In the next section, we discuss another, direct but rather involved proof that the measure at all $\lambda_j=0$ (\ref{me1}) is correct.

\section{Ward identities}

In this section, we consider the case of all $\lambda_j=0$, when the Ward identities for the matrix model partition function and its $\beta$-deformation are much simpler (see \cite{MMsc}). We derive the Ward identities for the partition function of model (\ref{PartFuncLambda}) in this case, and derive from them the $W$-representation following \cite{MMMR}. It coincides with the $W$-representation for the WLZZ models, and proves that our formula (\ref{PartFuncLambda}) reproduces (\ref{WLZZ}) at all $\lambda_j=0$. In fact, as we explained in sec.\ref{gen}, the claim at all $\lambda_j=0$ is enough in order to guarantee the proof at generic $\lambda_j$. Note that for the derivation of Ward identities the choice of integration contours is inessential, hence, we do not specify in this section that some of integrations run over the real axis and some other, over the imaginary ones.

\subsection{Matrix model: $\beta=1$ case}

To begin with, we consider the case of $\beta=1$. In this case, the multiple integral (\ref{PartFuncLambda}) is equivalent to the two-matrix model.

For the sake of simplicity, consider the Hermitian two-matrix model with potential of degree $s+1$:
\be\label{2mm}
Z_{s+1}(p)=\int [dXdY]\exp\left({1\over s+1}\Tr X^{s+1}-\Tr XY+\Tr V(Y)\right):=\int [dY]\exp\left(\Tr V(Y)\right)F(Y)
\ee
where $V(y)=\sum_ky^kp_k/k$. The function
\be
F(Y)=\int [dX]\exp\left({1\over s+1}\Tr X^{s+1}-\Tr XY\right)
\ee
satisfies the Ward identity
\be\label{WIF}
\Tr\left(Y^n{\p^s\over\p Y^s}-(-1)^sY^{n+1}\right)F(Y)=0,\ \ \ \ \ \ n\ge 0
\ee
and the matrix model (\ref{2mm}) is understood as a formal power series in variables $p_k$.

Then,
\be
(n+1){\p Z_{s+1}(p)\over\p p_{n+1}}=\int [dY]\exp\left(\Tr V(Y)\right)\Tr\Big( Y^{n+1}\Big)F(Y)=\nn\\
=(-1)^{-s}\int [dY]\exp\left(\Tr V(Y)\right)\Tr\left( Y^n{\p^s\over\p Y^s}\right)F(Y)=
\int [dY]F(Y)\Tr\left({\p^s\over\p Y^s}Y^n\right)\exp\left(\Tr V(Y)\right)
\ee
Now using the definition
\be
{\widetilde W}^{(s+1)}_{n-s}\exp\left(\Tr V(Y)\right)=\Tr\left({\p^s\over\p Y^s}Y^n\right)\exp\left(\Tr V(Y)\right)
\ee
where ${\widetilde W}^{(p+1)}_n$ is a differential operator in variables $p_k$,
one obtains the Ward identity
\be\label{nb}
(n+1){\p Z_{s+1}(p)\over\p p_{n+1}}={\widetilde W}^{(s+1)}_{n-s}Z_{s+1}(p)
\ee
In fact, the operator ${\widetilde W}^{(p+1)}_n$ is the corresponding element of the ${\widetilde W}^{(-,p+1)}$-algebra, however, as soon as we consider in the paper only the operators of this kind (and their $\beta$-deformations), we omit the superscript ``-".

Equation (\ref{nb}) is immediately solved \cite{MMMR}:
\be
Z_s(p)=e^{{1\over s}\hat H_s(p)}\cdot 1
\ee
with
\be
\hat H_s=\sum_kp_k{\widetilde W}^{(s)}_{k-s}
\ee
which is nothing but formula (\ref{1}).

Formula (\ref{nb}) is easily generalized to the generic potential of $X$ in (\ref{2mm}) as in (\ref{im}): the Ward identities in this case are
\be
(n+1){\p Z(N;p,0,g)\over\p p_{n+1}}=\sum_kg_k{\widetilde W}^{(k)}_{n+1-k}Z(N;p,0,g)
\ee
and
\be
Z(N;p,0,g)=e^{\sum_k{g_k\over k}\hat H_k(p)}\cdot 1
\ee
since all $\hat H_k$ are commuting.

\subsection{$\beta$-deformation}

As we demonstrated in \cite{MMCal,MMMP1,MMMP2}, the matrix realization of the elements of the $W_{1+\infty}$ algebra within the wedge, i.e. the area bounded by the rational Calogero Hamiltonians can be immediately $\beta$-deformed by replacing matrices $Y$ with their eigenvalues $y_i$, traces with sums over eigenvalues, and matrix derivatives with the Dunkl operators $\hat D_{j,y}$:
\be
\hat D_{j,y}={\p\over\p y_j}+\beta\sum_{k\ne j}{1\over y_j-y_k}(1-P_{jk})
\ee
where $P_{jk}$ is the operator permuting $y_j$ and $y_k$. As soon as the $\beta$-HCIZ function is an eigenfunction of the Calogero Hamiltonians, it is natural that the counterpart of (\ref{WIF}), in the case of $\beta$-deformation, is given by the equation (we prove it in the next subsection)
\be\label{main}
\boxed{
\left(\hat D_{j,y}^s-(-\beta)^sy_j\right)F_\beta(y)=0,\ \ \ \ \ \forall j
}
\ee
where
\be
F_\beta(y)=\int [dx]\Delta(x)^{2\beta}\exp\left({\beta\over s+1}\sum_j x_j^{s+1}\right)I^\beta(x,y)
\ee
and, at $\beta=1$,
\be\label{Fb}
F(y)=\int [dX]\exp\left({1\over s+1}\Tr X^{s+1}-\Tr XY\right)=\int [dx]{\Delta(x)\over\Delta(y)}\exp\left({1\over s+1}\sum_j x_j^{s+1}-\sum_j x_jy_j\right)
\ee
In fact, this realization of the $\beta$-deformation works only for operators acting on symmetric functions of the eigenvalues $y_i$, but $F(y)$ is just such a function.

The counterpart of the two-matrix model (\ref{2mm}) at all $\lambda_j=0$ and monomial potential in $X$, in accordance with that we discussed in sec.3 looks now (this two $\beta$-ensemble model was also considered in \cite{Ey1} along with the corresponding loop equations)
\be\label{zsb}
Z_{s+1}^{(\beta)}(p)=\int [dy]\left(\sum_j y_j^n\Delta(y)^{2\beta}\right)\exp\left(\beta\sum_k V(y_k)\right)F_\beta(y)
\ee
Then, one obtains the Ward identity
\be\label{W}
\boxed{
{(n+1)\over\beta}{\p Z_{s+1}^{(\beta)}(p)\over\p p_{n+1}}=(-\beta)^{-s}\int [dy]\left(\sum_j y_j^n\Delta(y)^{2\beta}\right)\exp\left(\beta\sum_k V(y_k)\right)\hat D_{j,y}^sF_\beta(y)
}
\ee
Again, using the integration by parts, one can reduce the r.h.s. of this expression to action of a differential operator ${\widetilde W}^{(s+1,\beta)}_{n-s}$ in variables $p_k$ to $\exp\left(\beta\sum_j V(y_j)\right)$, and obtain finally
\be\label{WIb}
\boxed{
\beta^{s-1}(n+1){\p Z_{s+1}^{(\beta)}(p)\over\p p_{n+1}}={\widetilde W}^{(s+1,\beta)}_{n-s}Z_{s+1}^{(\beta)}(p)
}
\ee
Examples of operators ${\widetilde W}^{(s+1,\beta)}_{n}$ can be found in sec.\ref{We}.

This equation immediately implies that (see formula (\ref{1b}))
\be
Z_s^{(\beta)}(p)=e^{{\beta^{1-s}\over s}\hat H_s(p)}\cdot 1
\ee
where
\be\label{HsW}
\hat H_s=\beta\sum_kp_k{\widetilde W}^{(s,\beta)}_{k-s}
\ee

These formulas are again easily generalized to the generic potential of $X$ in (\ref{W}) as in (\ref{imJ}): the Ward identities in this case are
\be
(n+1){\p Z^{(\beta)}(N;p,0,g)\over\p p_{n+1}}=\sum_k\beta^{2-k}g_k{\widetilde W}^{(k,\beta)}_{n+1-k}Z^{(\beta)}(N;p,0,g)
\ee
and
\be
Z^{(\beta)}(N;p,0,g)=e^{\sum_k{\beta^{1-k}\over k}g_k\hat H_k(p)}\cdot 1
\ee
since all $\hat H_k$ are again commuting.

\subsection{Proof of eqn.(\ref{main})}

First of all note that, in the case of $N=1$,
\be
F_\beta=\int dx\exp\left({\beta\over s+1}x^{s+1}-\beta xy\right)
\ee
and formula (\ref{main}) is evident.

In order to prove (\ref{main}), we need two identities. The first one is the identity for the $\beta$-HCIZ integral
\be\label{WIIZ}
\boxed{
\sum_j \Big(-\beta^{-1}\Big)^my_j^n \hat{D}_{j, y}^m I^\beta(x,-y)= \sum_j \Big(-\beta^{-1}\Big)^nx_j^m \hat {D}^n_{j, x} I^\beta(x,-y)
}
\ee
or
  \begin{align} \label{eq:iz-diff-duality}
    \sum_j y_j^n \hat{\tilde D}_{j, y}^m I^\beta(x,-y) = \sum_j x_j^m \hat {\tilde D}^n_{j, x} I^\beta(x,-y)
\end{align}
where we denoted for the sake of brevity $ \hat{\tilde D}:=-\beta^{-1}\hat D$. This identity is checked for various values of $m$ and $n$. What is important, it is valid for any $\beta$, and not obligatory natural, since it is checked for the definition of the $\beta$-HCIZ integral as a power series (\ref{IZJack}). Note that (\ref{WIIZ}) is evident for $N=1$, and for $m=0$, in the latter case being just a claim that the $\beta$-HCIZ integral is an eigenfunction of the rational Calogero Hamiltonians \cite[Eq.(198)]{MMMP1} (which are just $\sum_j \hat {\tilde D}^n_{j, x}$, \cite[Eq.(79)]{MMMP1}).

The second identity is
\be\label{si}
\sum_{i,j;i\ne j}{1\over y_i-y_j}P_{ij}f_i(y)=\sum_{i,j;i\ne j}{1\over y_i-y_j}f_j(y)=-\sum_{i,j;i\ne j}{1\over y_i-y_j}f_i(y)
\ee
Then, one can write
\be
0=\int [dx]\sum_j{\p\over\p x_j}\left[\Delta^{2\beta}(x)\exp\left({\beta\over s+1}\sum_kx_k^{s+1}\right)\hat {\tilde D}^n_{j, x} I^\beta(x,-y)\right]=\nn\\
=\int [dx]\Delta^{2\beta}(x)\exp\left({\beta\over s+1}\sum_kx_k^{s+1}\right)\sum_j\left[2\beta\sum_{k\ne j}
{1\over x_j-x_k}+\beta x_j^s+{\p\over\p x_j}\right] \hat {\tilde D}^n_{j, x} I^\beta(x,-y)\stackrel{(\ref{si})}{=}\nn\\
=\int [dx]\Delta^{2\beta}(x)\exp\left({\beta\over s+1}\sum_kx_k^{s+1}\right)\sum_j\left[\beta x_j^s+\hat D_{j,x}\right] \hat {\tilde D}^n_{j, x} I^\beta(x,-y)=\nn\\
=\beta\int [dx]\Delta^{2\beta}(x)\exp\left({\beta\over s+1}\sum_kx_k^{s+1}\right)\sum_j\left[x_j^s
- \hat {\tilde D}_{j, x}\right] \hat {\tilde D}^n_{j, x} I^\beta(x,-y)\stackrel{(\ref{eq:iz-diff-duality})}{=}\nn\\
=\beta\int [dx]\Delta^{2\beta}(x)\exp\left({\beta\over s+1}\sum_kx_k^{s+1}\right)\sum_j\left[y_j^n\hat {\tilde D}^s_{j, y}
- y_j^{n+1}\right] I^\beta(x,-y)=\beta\sum_j\left[y_j^n\hat {\tilde D}^s_{j, x}-y_j^{n+1}\right] F_\beta(y)
\ee
which is nothing but formula (\ref{main}).

One can find in the Appendix some illustrations of how formula (\ref{main}) works in various particular cases.

\subsection{Examples of operators ${\widetilde W}^{(s+1,\beta)}_{n}$\label{We}}

Consider the first few examples of the operators ${\widetilde W}^{(s+1,\beta)}_{n}$.

\paragraph{$\boxed{{\widetilde W}^{(2,\beta)}_{n}}$} We start with the Gaussian potential $s=1$.
In this case, one writes at $n>1$
\be
-\int [dy]\left(\sum_j y_j^n\Delta(y)^{2\beta}\right)\exp\left(\beta\sum_k V(y_k)\right)\hat D_{j,y}F_\beta(y)=
\int [dy]F_\beta(y)\sum_j {\p\over\p y_j}\left[y_j^n\Delta(y)^{2\beta}\exp\left(\beta\sum_k V(y_k)\right)\right]=\nn\\
=\int [dy]F_\beta(y)\Delta(y)^{2\beta}\sum_j
\left[ny_j^{n-1}+2\beta\sum_{j,k;j\ne k}{y_j^n\over y_j-y_k}+\beta\sum_kp_ky_j^{k+n-1}\right]\exp\left(\beta\sum_k V(y_k)\right)=\nn\\
=\int [dy]F_\beta(y)\Delta(y)^{2\beta}\left[{(1-\beta)n(n-1)\over\beta}{\p\over\p p_{n-1}}+\beta^{-1}\sum_{a,b>0}^{a+b=n-1}ab{\p^2\over \p p_a\p p_b}+2N(n-1){\p\over\p p_{n-1}}+\right.\nn\\
\left.+\sum_k(k+n-1)p_k{\p \over \p p_{k+n-1}}\right]\exp\left(\beta\sum_k V(y_k)\right)=
{\widetilde W}^{(2,\beta)}_{n-1}Z
\ee
where we used that
\be
2\sum_{j,k;j\ne k}{y_j^n\over y_j-y_k}=\sum_{j,k;j\ne k}{y_j^n-y_k^n\over y_j-y_k}=\sum_{a,b=1}^{a+b=n-1}\left(
\sum_jy_j^a\right)\left(\sum_ky_k^b\right)+(2N-n)\sum_jy_j^{n-1}
\ee
At $n=1$, one obtains
\be
-\int [dy]\left(\sum_j y_j\Delta(y)^{2\beta}\right)\exp\left(\beta\sum_k V(y_k)\right)\hat D_{j,y}F_\beta(y)=\nn\\
=\int [dy]F_\beta(y)\Delta(y)^{2\beta}\sum_j
\left[1+2\beta\sum_{j,k;j\ne k}{y_j\over y_j-y_k}+\beta\sum_kp_ky_j^{k}\right]\exp\left(\beta\sum_k V(y_k)\right)=\nn\\
=\int [dy]F_\beta(y)\Delta(y)^{2\beta}\left[(1-\beta)N+\beta N^2+\sum_kkp_k{\p \over \p p_k}\right]\exp\left(\beta\sum_k V(y_k)\right)=
{\widetilde W}^{(2,\beta)}_0Z
\ee
At last, at $n=0$, one obtains
\be
-\int [dy]\left(\sum_j \Delta(y)^{2\beta}\right)\exp\left(\beta\sum_k V(y_k)\right)\hat D_{j,y}F_\beta(y)=\nn\\
=\int [dy]F_\beta(y)\Delta(y)^{2\beta}\sum_j
\left[2\beta\underbrace{\sum_{j,k;j\ne k}{1\over y_j-y_k}}_{=0}+\beta\sum_kp_ky_j^{k-1}\right]\exp\left(\beta\sum_k V(y_k)\right)=\nn\\
=\int [dy]F_\beta(y)\Delta(y)^{2\beta}\left[\beta Np_1+\sum_k(k-1)p_k{\p \over \p p_{k-1}}\right]\exp\left(\beta\sum_k V(y_k)\right)=
{\widetilde W}^{(2,\beta)}_{-1}Z
\ee
Thus, one finally obtains
\be\label{W2b}
\boxed{
\begin{array}{c}{\widetilde W}^{(2,\beta)}_{n}=\sum_k(k+n)p_k{\p \over \p p_{k+n}}+\beta^{-1}\sum_{a,b>0}^{a+b=n}ab{\p^2\over \p p_a\p p_b}
+2Nn{\p\over\p p_{n}}+\cr
\cr
+{(1-\beta)n(n+1)\over\beta}{\p\over\p p_{n}}
+(1-\beta)N\delta_{n,0}+\beta N^2\delta_{n,0}+\beta Np_1\delta_{n,-1}
\end{array}
}
\ee
which is, at $\beta=1$, the standard expression for the Virasoro algebra constraints for the Gaussian Hermitian one matrix model \cite{Vir1,Vir2,Vir3,Vir4}.

\paragraph{$\boxed{{\widetilde W}^{(3,\beta)}_{n}}$} Similarly, one obtains at $p=2$
\be\label{W3b}
\boxed{
\begin{array}{c}{\widetilde W}^{(3,\beta)}_{n}=n(n+1)(n+2) \frac{ 2-3\beta+2\beta^2 }{ 2\beta } \frac{\partial}{\partial p_{n}} + (1-\beta) \sum_{k} p_k (k+3+2n)(n+k) \frac{\partial}{\partial p_{n+k}} +\cr
\cr
        +2\sum_{k}p_k \sum_{a=0}^{p+k-2} a(n+k-a) \frac{\partial^2}{\partial p_{a} \partial p_{n+k-a}}-\sum_{k}p_k \sum_{a=0}^{k-2} a(n+k-a) \frac{\partial^2}{\partial p_{a} \partial p_{n+k-a}} +\cr
        \cr
        +\frac{1}{\beta} \sum_{a=0}^{n} \left[ \frac{3}{2}(n+2) -n \beta - \beta (a+3) \right] \frac{\partial^2}{\partial p_{a} \partial p_{n-a}}+\beta \sum_{k,l} p_k p_l \frac{\partial}{\partial p_{n+k+l}}+\frac{1}{\beta} \sum_{a+b+c=n} abc \frac{\partial^3}{\partial p_{a}\partial p_{b}\partial p_{c}} +\cr
        \cr
        + n(n+2)\frac{ 2-3\beta+2\beta^2 }{ 2\beta } N \delta_{n,0}+2\beta(1-\beta)N p_1 \delta_{n,-1}+\beta N(\beta p_1^2+ (1-\beta) p_2)\delta_{n,-2}
\end{array}
}
\ee

\subsection{Connection to WLZZ models}

Now one checks that
\begin{itemize}
\item the partition function
\be
Z_\beta^{(s)}=\sum_R||J_R||^{-1}\xi_R^\beta(N)J_R\{p_k\}J_R\{\delta_{k,s+1}\}
\ee
solves eqn.(\ref{WIb}) with ${\widetilde W}^{(2,\beta)}_{n}$ as in (\ref{W2b}) at $s=1$ and with ${\widetilde W}^{(3,\beta)}_{n}$ as in (\ref{W3b}) at $s=2$.
\item the constructed ${\widetilde W}^{(\beta)}$-operators lead to the Hamiltonians $\hat H_2$ and $\hat H_3$ of sec.2.1 upon using formula (\ref{HsW})
\be
\hat H_s=\beta\sum_kp_k{\widetilde W}^{(s,\beta)}_{k-s}
\ee
\end{itemize}

Thus, indeed, our formula (\ref{PartFuncLambda}) reproduces (\ref{WLZZ}) at all $\lambda_j=0$.

\section{Concluding remarks}

In this paper, we considered a two $\beta$-ensemble system in the external field, and demonstrated that it describes the infinite series of the $\beta$-deformed WLZZ models. Our main claim is that the multiple integral (\ref{PartFuncLambda}) is equal to the partition function of the WLZZ models obtained via the $W$-representation (\ref{WLZZ}) and is given by the corresponding sums of Jack polynomials over all partitions.

Technically, our claim simply follows from the case without the external field, and this latter one is checked either using one of the representations of the $\beta$-HCIZ formula at $\beta\in\mathbb{N}$ (\ref{iizrec}), or via the Ward identities. We derived the Ward identities (\ref{WIb}) in this case basing on the identity (\ref{WIIZ}) for one $\beta$-ensemble in the external field
(\ref{Fb}), the identity itself being a corollary of a striking identity (\ref{WIIZ}) for the $\beta$-HCIZ integral. This latter identity arguably makes the $\beta$-HCIZ integral an indispensable tool for constructing even more complicated matrix models associated with the affine Yangian: a task for future developments.

As a by-product of constructing these Ward identities, we obtained a natural $\beta$-deformation of the notion of $\widetilde W$-algebra (\ref{W}), which allowed us to construct manifest examples of such algebras of spins two (\ref{W2b}) and three (\ref{W3b}).

It, however, remains a challenge to find an effective technique of evaluating $\widetilde W$-algebra generators: we know such technique only in the matrix model, i.e. $\beta=1$ case \cite{MMM91,GKMU,DMP}. Another problem is to construct $\widetilde W$-algebras associated with all commutative subalgebras of the affine Yangian corresponding to the integer rays \cite{MMMP1,MMMP2}, i.e. with Hamiltonians of generalizations of the rational Calogero system \cite{MMCal}. These $\widetilde W$-algebras would generalize the construction of \cite{DMP} at $\beta\ne 1$.

In fact, constructing the matrix/$\beta$-ensemble models that describe the WLZZ models associated with these generalizations of the rational Calogero system remains unsolved problem so far even at $\beta=1$, and this definitely deserves further investigation.

\section*{Acknowledgements}

We are grateful to A. Alexandrov, Ya. Drachev and A. Morozov for useful discussions. This work was supported by the Russian Science Foundation (Grant No.23-41-00049).

\section*{Appendix. Formula (\ref{main})}

In this Appendix, we demonstrate how formula (\ref{main}) works in simple examples. We will consider integrals like
\be
F(y)=\int [dx]{\Delta(x)\over\Delta(y)}\exp\left(-{1\over s+1}\sum_j x_j^{s+1}+i\sum_j x_jy_j\right)
\ee
and
\be
F_\beta(y)=\int [dx]\Delta(x)^{2\beta}\exp\left(-{\beta\over s+1}\sum_j x_j^{s+1}\right)I^\beta(x,iy)
\ee
to have all integrations over the real axis and for convergency of the Gaussian integral.

\paragraph{Gaussian potential at $\beta=1$.} The simplest check is the Gaussian potential, when the equation is
\be
\left(\hat D_j+y_j\right)F(y)=0,\ \ \ \ \ \forall j
\ee
In this case, the second term in the Dunkl operator does not effect the result. On the other hand, one obtains that
\be
F(y)\sim \exp\left(-{1\over 2}\sum_jy_j^2\right)
\ee
which certainly satisfies (\ref{main}).

\paragraph{Gaussian potential at generic $\beta$ and $N=2$.} In this case, \cite{Brezin,Ey}
\be\label{IZb2}
I^\beta(x,iy)={\exp\left(ix_1y_1+ix_2y_2\right)\over z^{2\beta}}Q_\beta(z)+{\exp\left(ix_1y_2+ix_2y_1\right)\over (-z)^{2\beta}}Q_\beta(-z)
\ee
where
\be
Q_\beta(x)=\sum_{k=0}{\Gamma (\beta+k)\over\Gamma(\beta-k)}{x^{\beta-k}\over 2^kk!}
\ee
and
\be
z=-{i\over 2}(x_1-x_2)(y_1-y_2)
\ee
Then, one again immediately obtains
\be
F(y)\sim \exp\left(-{1\over 2}\sum_{j=1}^2y_j^2\right)
\ee

\paragraph{Cubic potential at $\beta=1$.} Now let us consider a more complicated case of the cubic potential at $\beta=1$.
Let us denote
\be
\Big<\ldots\Big>_F=\int [dx]{\Delta(x)\over\Delta(y)}\exp\left(-{1\over 3}\sum_j x_j^3+i\sum_j x_jy_j\right)\ldots
\ee
Then, inserting the total derivatives ${\p\over\p x_j}$ to the integrand, one obtains the Ward identity:
\be
0=\left<-x_j^2+\sum_{k\ne j}{1\over x_j-x_k}+iy_j\right>_F
\ee

One can check the main identity (\ref{main}) directly from the matrix integral. We note that \cite{GKM}
\be
F(y)=\int [dx]{\Delta(x)\over\Delta(y)}\exp\left(-{1\over 3}\sum_j x_j^3+i\sum_j x_jy_j\right)=
\int [dx]{\det_{i,j}x_i^{j-1}\over\Delta(y)}\exp\left(-{1\over 3}\sum_j x_j^3+i\sum_j x_jy_j\right)=\nn\\
={\det_{i,j}f_i(y_j)\over\Delta(y)}={\det_{i,j}f^{(i-1)}(y_j)\over\Delta(y)}
\ee
where we denote the Airy integral moments
\be
f_i(y)=\int dxx^{i-1}\exp\left(-{1\over 3}x^3+ixy\right)
\ee
denote $f_1=F_1=f$, and $f^{(j)}(y)$ is the $j$-th derivative of $f$.

Now one notes that, for the Airy integral, one has
\be
f''=-iyf\nn\\
f'''=-iyf'-if
\ee
etc. Then, at $N=2$,
\be
F_2={f'(y_1)f(y_2)-f(y_1)f'(y_2)\over y_1-y_2}
\ee
and
\be
\sum_jy_j^n\hat D_j^2 F_2+i\sum y_j^{n+1}F_2=0
\ee
which is checked by a direct computation.

A similar calculation in the case of $N=3$ gives rise to the same Ward identity
\be
\sum_jy_j^n\hat D_i^2 F_3+i\sum y_j^{n+1}F_3=0
\ee
with $F_3$
\be
F_3={1\over\Delta}\det\left(
\begin{array}{ccc}
f(y_1)&f(y_2)&f(y_3)\cr
f'(y_1)&f'(y_2)&f'(y_3)\cr
f''(y_1)&f''(y_2)&f''(y_3)
\end{array}
\right)
\ee
One can similarly check it for higher $N$.

\end{document}